\documentclass[%
 aip,
amsmath,amssymb,
 reprint,%
]{revtex4-1}
\usepackage[version=3]{mhchem} 

\usepackage{natbib}
\usepackage{graphicx}
\usepackage{bm}
\usepackage{color,soul}
\usepackage{hyperref}
\usepackage{float}
\usepackage{csquotes}
\usepackage{bibentry}

\usepackage{color,soul}
\usepackage{hyperref}
\hypersetup{colorlinks=true, urlcolor=blue, citecolor=blue}

\makeatletter
\def\@email#1#2{%
 \endgroup
 \patchcmd{\titleblock@produce}
  {\frontmatter@RRAPformat}
  {\frontmatter@RRAPformat{\produce@RRAP{*#1\href{mailto:#2}{#2}}}\frontmatter@RRAPformat}
  {}{}
}%
\makeatother
\begin{document}

\title{ Feeble Metallicity and Robust Semiconducting Regime in Structurally Sensitive Ba(Pb, Sn)O$_3$ Alloys
}

\author{Ravi Kashikar}

\author{B. R. K. Nanda}
\email{nandab@iitm.ac.in}

\affiliation{
 Condensed Matter Theory and Computational Lab, Department of Physics,
Indian Institute of Technology Madras, Chennai - 36, India
}

\begin{center}
    \date{\today}
\end{center}

\begin{abstract}
Density functional calculations are carried out to study the symmetry and substitution-driven electronic phase transition in BaPb$_{1-x}$Sn$_x$O$_3$. Two end members BaSnO$_3$ and BaPbO$_3$, are found to be insulating and metallic, respectively. In the latter case, the metallicity arises with the presence of an electron pocket, formed by Pb-s dominated conduction band edge, and a hole pocket formed O-p dominated valence bands. While electron carriers are found to be highly mobile, the hole carriers are localized. Our study reveals that an insulating phase can be realized in the metallic cubic BaPbO$_3$ in three ways in order to explore optoelectronic properties. Firstly, by lowering the symmetry of the lattice to monoclinic through rotation and tilting of the PbO$_6$ octahedra. Secondly, by hydrostatic pressure, and thirdly by alloying with Sn substitution. The presence of soft phonon modes implies the plausibility of symmetry lowering structural transitions. Furthermore, unlike the earlier reports, we find that Sn substituted BaPbO$_3$ cannot exhibits topological insulator phase due to absence of the band inversion. 
\end{abstract}

\maketitle

Weakly correlated oxide perovskites of the form ABO$_3$, where A is an alkali or alkaline earth element and B = Sn, Pb, Bi, have shown diverse structural polymorphs and electronic properties, including metal-insulator transition, superconductivity, and nontrivial topological phases \cite{11,12,13,14,15,16,17,18}. Among these, the emerging superconducting properties, similar to that exhibited by cuprates, have been investigated extensively in their pristine form or with substitution at A and/or B sites \cite{21,22,23}. BaPb$_{1-x}$B$_x$O$_3$ is one such member in this family which exhibits Drude kind of metallic behaviour in its pristine form (x=0) and superconducting phase upon Bi (0$<$x$\leq$0.3) substitution at T$_c$ = 13 K \cite{30,31,32,33,34,35}. The solid solution also demonstrates the metal to insulator transition at x=0.35.

In recent years, bilayers and interfaces made out of the two end-members BaBiO$_3$ (BBO) and BaPbO$_3$ (BPO) have gained attention to realize two-dimensional superconductivity\cite{BPO_1,BPO_2,PNAS_BPO}. Furthermore, these are also explored as promising topological materials under pressure and chemical doping\cite{16,17}. The band structures reported in the literature show the existence of band inversion -a necessary criteria for topological insulating (TI) phase- in cubic BPO \cite{16}, and BBO \cite{16,17} between Bi (Pb)-s and -p dominated bands, occurs in both bonding and antibonding spectrum.  However,  these inverted orbital states exist around E$_F$-5 eV and E$_F$+5 eV, and shifting the Fermi level by 5 eV through excessive hole or electron doping is nearly impossible. Nevertheless, a very recent experimental study on Sn substituted BaPbO$_3$ thin films by Shiogai et al. have demonstrated orbital inversion between Pb-s dominated conduction band and O-p valence band at the Fermi level in the band spectrum \cite{BPO_sn_prb}. Thereby, the Sn substituted BPO is proposed as a promising class of topological compound. However, no further experimental and theoretical studies are available to establish the nontrivial band structure in this system.

The unusual properties exhibited by the members of this family are mainly driven by the  structural phase transition either due to temperature or due to substitution. From the structural point of view, BPO crystallizes in cubic, tetragonal, orthorhombic polymorphs in various temperature ranges, and all the experimental works have reported the cubic phase at high temperature\cite{struct1,struct2,struct3}. However, there exists a conflicting conclusion in these reports regarding low temperature phase. Ritter et al. and Ksepko et al. have identified the monoclinic phase with the space group I2/m at room temperature (RT) as well as at low temperature \cite{Ritter1989, Ksepko2018}. On the other hand, Mossa et al. have characterized the structure of BPO as orthorhombic phase (Imma) at RT and monoclinic phase (I2/m) at low temperature \cite{struct2}. As far as the electronic structure of BPO is concerned, the earlier studies are restricted to cubic phase and studies have concluded the metallic nature of cubic BPO \cite{Mathias, APW_cubic}. However, the x-ray absorption spectrometry (XAS) results concluded the BPO as an insulator \cite{Expt1,Expt2,Expt3}. Furthermore, in these studies it is emphasized that discrepancies for BPO must be addressed  through the proper electronic structure studies in various crystal phases. Therefore, we believe a detailed study has to be carried out to establish the correlation between structural phase transition with phonon and electronic band structure.

In the present work, we analyze both electronic, and phonon band structures of various polymorphs of BPO and Sn substituted BPO alloys by employing density functional theory to address the aforementioned issues; specifically, (I) low temperature crystal structure phase, (II) robustness of the metallic nature in BPO and forces that drive the metal-insulator transitions, (III) verifying the probable presence of band inversion in the Sn substituted BPO alloys as claimed experimentally.

 BPO exhibits various structural phases in different temperature ranges. The corresponding experimental details are listed in Table \ref{T1}. In the tetragonal phase, the adjacent octahedra are rotated in the $ab$ plane. Whereas in the orthorhombic phase, along with the rotation, a tilt is observed along the $c$ axis. The monoclinic phase forms with the minor distortion in the orthorhombic phase, i.e., with the axial angle $\gamma \neq 90^\circ$ \cite{struct1,struct2,struct3}.
 
\begin{table}
\footnotesize
    \centering
    \caption{ Lattice parameters, space group and temperature range of crystallization of BPO, BPO polymorphs and cubic BSO structures \cite{struct1,struct2,struct3,BaSno_struct}. }
    \begin{tabular}{cccc}
    \hline
    \hline
    Compound&Space Group&Lattice & Octahedral  \\
          &Temperature (K) &Parameter (\AA) & Rotation Angle\\
    \hline
BaPbO$_3$&Pm-3m&a=4.29& $\theta_{ab}$=$\theta_c$ = 180$^{\circ}$  \\
&($>$673K)&&  \\
&I4/mcm& a=b=6.06& $\theta_{ab}$= 169.9$^{\circ}$\\
&(573-673K)&c=8.58&$\theta_c$ = 180$^{\circ}$  \\
&Imma&a=6.06& $\theta_{ab}$ = 167.9$^{\circ}$ \\
&($<$573K)&b=6.02,c=8.5& $\theta_c$ = 162.86$^{\circ}$ \\
&C2/m&a = 8.53, b= 6.07&$\theta_{ab}$ = 164.15$^\circ$\\
&$<$ 20K&c = 8.48, $\gamma$ = 135.2$^\circ$ & $\theta_{c}$ = 161.27$^\circ$\\
BaSnO$_3$&Pm-3m&a=4.12& $\theta_{ab}$=$\theta_c$ = 180$^{\circ}$\\

\hline
\hline
    \end{tabular}
\label{T1}
\end{table}

In Fig. \ref{fig1}, we have shown the orbital resolved band structure along with the band edges parity in terms of Koster notations. The electronic structure calculations are carried out using full potential based linearized augumneted plane wave (FP-LAPW) method as implemented in WIEN2k \cite{Blaha}. Perdew-Burke-Ernzerhof (PBE) exchange-correlation functional within the generalized gradient approximation (GGA) along with the modified Becke-Johnson (mBJ) exchange potential is employed for accurate bandgap estimation \cite{Perdew136406,mbj-1,mbj-2}. The cubic band structure presented in the Fig. \ref{fig1}a is the outcome of strong covalent interaction between Pb-\{s, p\}-O-p states which gives rise to bonding and antibonding bands along with nonbonding bands.  Both bonding and antibonding bands consist of four dispersive bands dominated with hybridized Pb-s and O-p molecular orbitals. The valence electron count in BPO turns out to be 18 such that the Fermi level lies between antibonding and nonbonding bands. This is similar to what has been observed in ABiO$_3$, where Bi-\{s, p\} and O-p covalent interactions governed antibonding and bonding spectrum.  However, in BPO, there is a significant contribution coming from Ba-d states in the unoccupied antibonding band spectrum. This is due to the fact that the s-p antibonding bands undergo rehybridization with unoccupied Ba-d orbitals.  The orbital of oxygen ligand exhibits symmetry dependent linear combinations at different high symmetry points so as to bond with the central Pb atom. The point group analysis of high symmetry $k$-points in the cubic Brillouin zone infers that $R$ and $\Gamma$ possess O$_h$ symmetry, whereas M and X have D$_{4h}$ symmetry. In the present case, for O$_h$ point group, nine oxygen p orbitals produce four sets of basis, namely, T$_{2u}$, T$_{2g}$, e$_g$ and A$_{1g}$. Among these four sets, A$_{1g}$ and T$_{2u}$ undergo hybridization with Pb-s and -p orbitals which has the symmetry adopted basis set as A$_{1g}$ and T$_{2u}$. The linear combination of this basis set forms four $\sigma$ bands, four $\pi$ bands and five nonbonding bands, and this is schematically illustrated in Fig. S1 of the supplementary information.

 The lowermost band in the antibonding spectrum, which is of Pb-s character, having larger bandwidth ($\sim$ 7 eV) and crosses the Fermi level at $\Gamma$ point (with the parity $\Gamma_{6+}$) to create the metallic nature of BPO. It rises upward as we move along $\Gamma$-X-M-R. In the presence of strong next neighbour Pb-Pb interactions, at R, the band inverts its character as reflected in the pair of R$_{6+}$ and R$_{6-}$. The band inversion changes Pb-s character to Ba-d+Pb-p at R. Earlier, Li et al. have claimed that the band inversion occurs between Pb-s and Pb-p \cite{14}. However,  we found that Pb-p orbital weight is $\sim$ 0.33 and that of Ba-d is $\sim$0.65. Such band inversion exists in the absence of SOC and is driven by strong second neighbour interactions \cite{17}. Nevertheless, these band inversion occurs above the Fermi level. 
\begin{figure}[h]
\centering
\includegraphics[angle=0.0,origin=c,height=10.5cm,width=8.5cm]{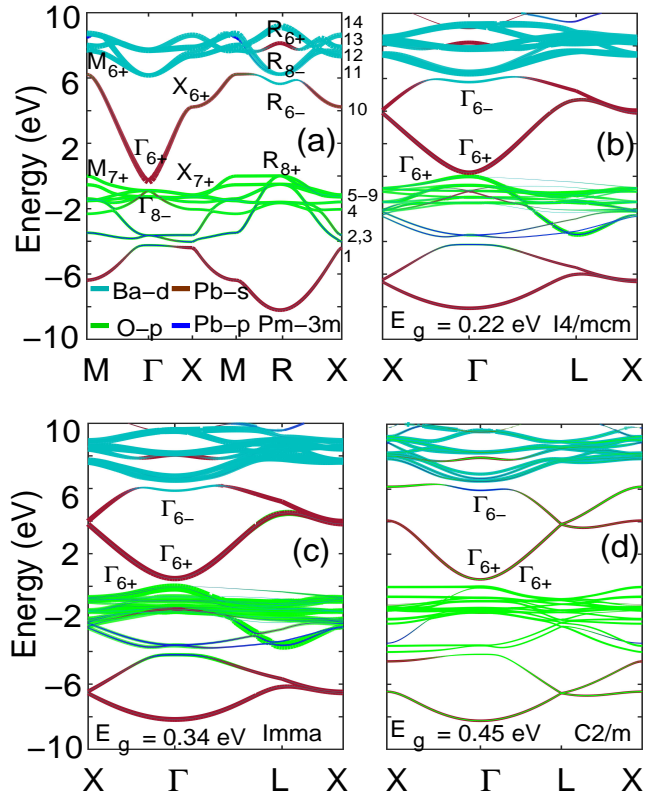}
\caption{ Orbital resolved band structure of cubic (a), tetragonal (b), orthorhmobic (c), and monoclinic (d) polymoprphs of BPO. The cubic BPO exhibits metallic phase whereas the lower crystal polymorphs show insulating phase due to octahedral rotations. The parity of the band edges at different high symmetry $k$-points are represented in the Koster notations. In cubic phase the bands are numbered so that it can be related to the molecular orbital picture and Table S1 of the SI. }
\label{fig1}
\end{figure}

 The band structures presented in Fig. \ref{fig1}(b-d) for lower symmetric crystal polymorphs of BPO reveal that there exists metal to insulator transition (MIT) as we move from cubic to tetragonal to orthorhombic phase. The phase transition introduces the semiconducting nature with a bandgap of 0.22 eV, 0.34 eV, and 0.45 eV in tetragonal, orthorhombic, and monoclinic phases, respectively. The rotation and tilting of the octahedra are found to be driving forces for MIT.  We observed the continuous decrease of bandgap with in-plane octahedral rotation ($\theta_{ab}$) (shown in Fig. S2 of the SI). The system enters metallic phase  around $\theta_{ab}$ = 175$^{\circ}$. The rotation weakens the covalent hybridization, leading to narrowing of the bandwidth, which results in bandgap formation. As the $sp$ based perovskites have also been investigated from an optoelectronic perspective, it is prudent here to examine the possibility of optical transitions.  The valence band maximum (VBM) and conduction band minimum (CBM) have the same parity and hence forbid optical transitions among them. However, the transition between the second most conduction band with parity ($\Gamma_{6-}$) and the VBM ($\Gamma_{6+}$) can occur.

\begin{figure}[h]
\centering
\includegraphics[angle=0.0,origin=c,height=4.5cm,width=8.5cm]{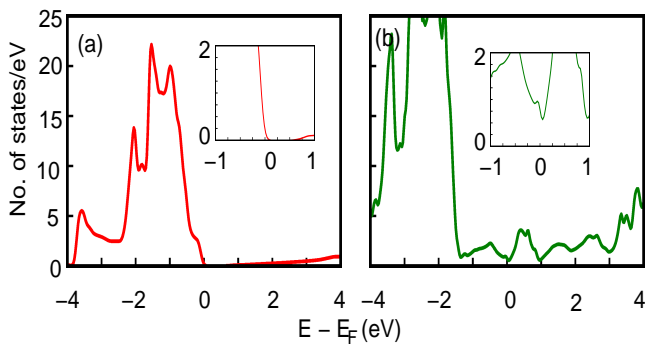}
\caption{ (a) Density of states of orthorhombic BaPbO$_3$ and (b) oxygen vacated orthorhombic BaPbO$_{2.916}$. BPO containing oxygen vacancies exhibits metallic nature.   }
\label{fig2}
\end{figure}
The electronic structure of orthorhombic BPO opens up the gap due to octahedral rotations. However, experimentally, metallic transport phenomena have been discussed. One of the possible reasons for this is free carriers from the incompletely filled $d$ orbitals as suggested by Hesigh and Fu \cite{HSIEH1992289}. BaPbO$_3$ is made up of BaO and PbO$_2$ layers. The charge on Pb in PbO$_2$ layer is {+4}, and thus has the electronic configuration [Xe]4f$^{14}$6s$^1$5d$^9$ instead usual [Xe]4f$^{14}$5d$^{10}$.  Another possibility is the existence of a small fraction of Pb atoms at Ba sites due to similar ionic radii, enhancing the carrier concentration and thus metallic behavior of BPO.   Groot et al. from the experimental observations concluded that the metallicity arises from the presence of oxygen vacancies in the crystal systems, where they observe a change in the resistivity with vacancy concentration \cite{Expt2}. The same observation has been made by Namtame et al. \cite{Expt1}. In fact Franz et al. have synthesized the BaPbO$_3$ with chemical composition BaPbO$_{2.74}$ and BaPbO$_{2.82}$ \cite{vacancy}. The authors also observed that the BaPbO$_3$ can only be synthesized with vacancies. Therefore, here we examine the effect of vacancies on the electronic structure.The  electronic structure of orthorhombic BaPbO$_{3-\delta}$, for $\delta \sim 0.08$, and corresponding DOS (shown in Fig.\ref{fig2}) confirms the metallic nature. We also calculated the electrical conductivity for both cubic and orthorhombic phases using the Boltzmann transport equation as implemented in BoltzTrap \cite{BT} and is presented in Fig. S3 of the SI. We found that there is minimal difference in the magnitude of conductivity ($\sigma/\tau$) in the conduction band region.

\begin{figure}[h]
\centering
\includegraphics[angle=0.0,origin=c,height=11.0cm,width=8.9cm]{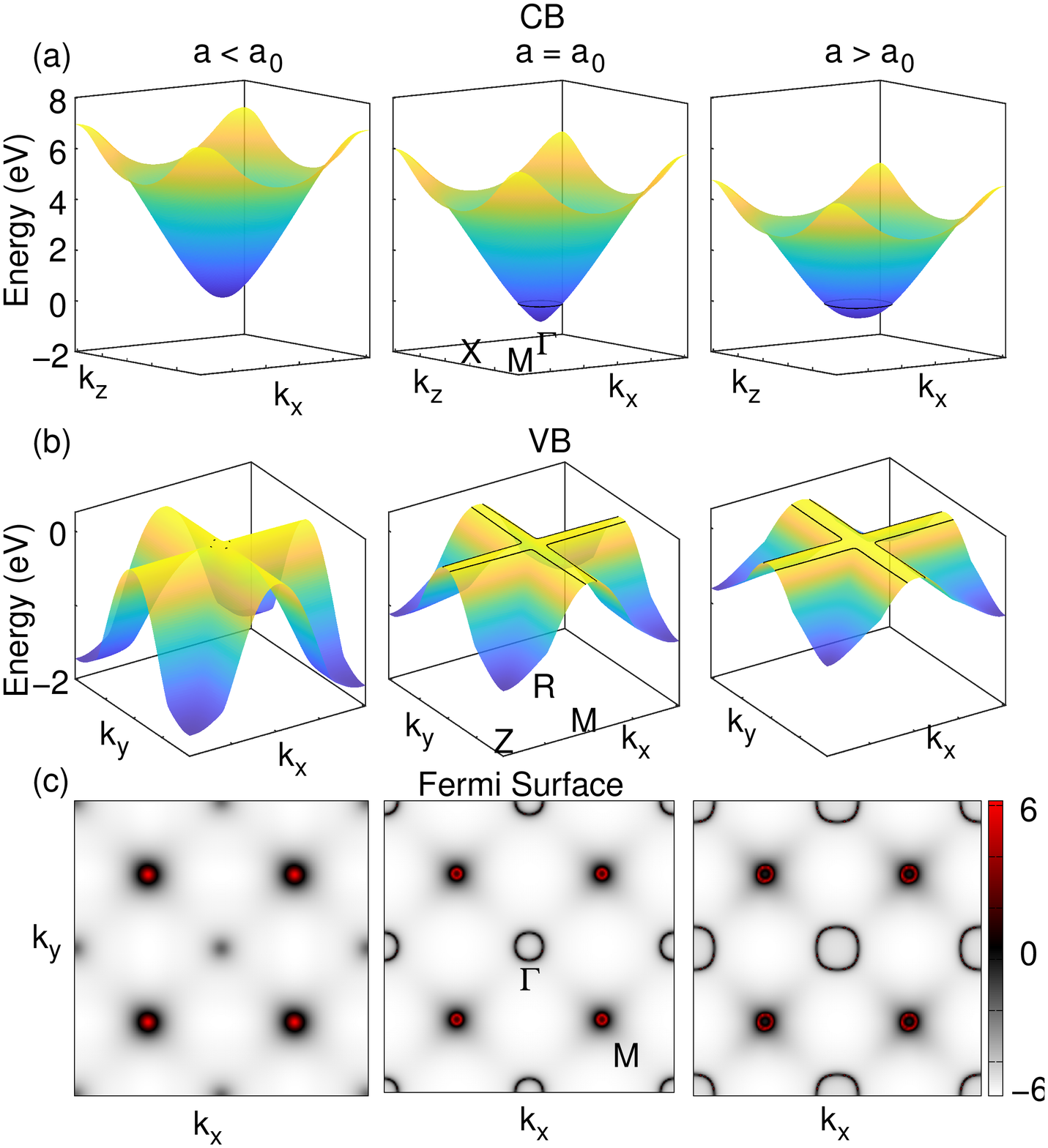}
\caption{ Three dimensional plot of conduction and valence bands in (a) $k_x$-$k_z$ (for $k_y$ = 0) and (b) $k_x$-$k_y$ (for $k_z$ = $\frac{\pi}{a}$) planes, respectively, for various cubic BPO cell parameters. (c) The corresponding Fermi surface in k$_x$-k$_y$ (for k$_z$ = 0) plane obtained from the Wannier function method. We observe the size of electron and hole pocket increases with lattice expansion.}
\label{fig3}
\end{figure}

We further examine the robustness of the metallic phase of cubic BPO by applying uniform compression and expansion on the cubic lattice. The 3D representation of conduction band which crosses the Fermi level at $\Gamma$ (see Fig. \ref{fig3}a), is shown in the k$_x$-k$_z$ plane in the upper panel of Fig. \ref{fig3}. Similarly, the flat valence band edge that lies on the Fermi level, centered around R, is shown in the middle panel of Fig. \ref{fig3}. With compression ($a<a_0$), the conduction band edge pushes above and becomes more dispersive, and as a consequence, an electron pocket formed in the equilibrium structure vanishes. In the case of expansion, the bandwidth narrows down to increase the area of an electron pocket, though the carriers are less mobile. In the valence band edge, the flatness decreases with compression, and as the state lies below the Fermi level, a clear bandgap emerges. With the expansion, the flatness increases as a result, heavy holes lie on the Fermi level. The Fermi surface plot in the lower panel of Fig. \ref{fig3}, as calculated using the Wannier formalism, further substantiates the disappearance of electron pocket with compression. The red circles illustrate hole pocket along $\Gamma$-M-$\Gamma$ (see Fig. \ref{fig1}a) which become larger with expansion.

 To study the reason behind the temperature-driven phase transitions in BPO, we have calculated the phonon band structure on the DFT-GGA optimized unit cells (see supplementary information for computational details), and the results are shown in Fig. \ref{fig4}.   The cubic phase phonon band structure shown in Fig. \ref{fig4}a exhibits triply degenerate and nondegenerate soft modes, identified through negative frequencies, at R and M high symmetry points, respectively. In the irreducible representation scheme, they are denoted as $R_{4+}$ and $M_{3+}$ for the point group O$_h$, and D$_{4h}$, respectively \cite{Rondinelli_phonon}. The triply degenerate modes at R are further denoted as $R_a$, $R_b$ and $R_c$. From the calculation of partial phonon density of states, we find that these modes are contributed from oxygen atoms and suggest that the displacement of the anions will reduce the instability. The type of displacement due to soft modes at M and R are depicted in Fig. \ref{fig4}a. The soft modes at M are weakened through the rotations of octahedra in the $ab$ plane leading to the tetragonal phase. This phase is characterised by out of phase rotations of octahedra around the $c$ axis with Glazer notations $a^0a^0c^-$ \cite{struct1,struct2,struct3}.  However, the phonon bands of this tetragonal phase, shown in Fig. S4 of the SI, infer that the soft modes still exist though with reduced strength. In case of orthorhombic phase also, we notice the presence of soft modes, albeit much depleted. However, we observe single soft mode in the monoclinic phonon band structure as shown in Fig. \ref{fig4}b.

\begin{figure}[h]
\centering
\includegraphics[angle=0.0,origin=c,height=9.0 cm,width=9.cm]{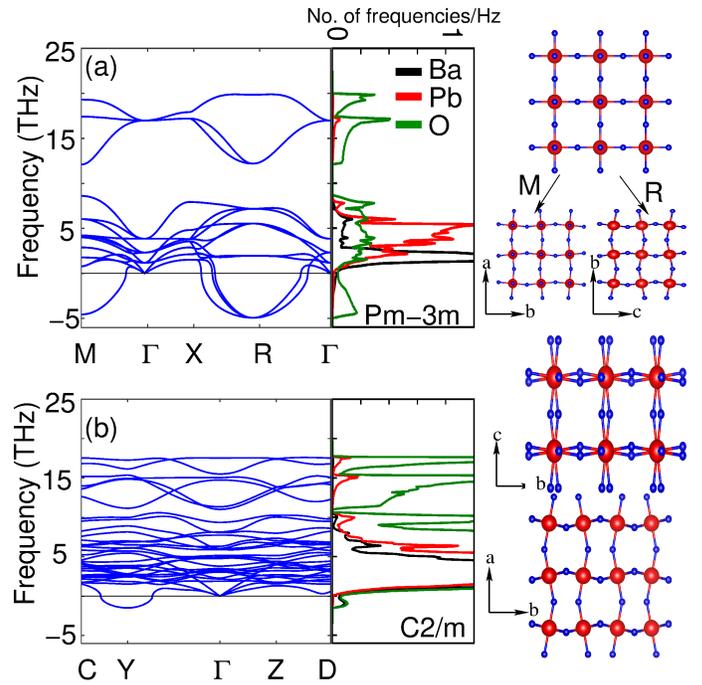}
\caption{Phonon band dispersion for (a) cubic  (b) monoclinic  BPO obtained from density functional perturbation theory. The soft modes in the cubic phase band dispersions infer that the structure undergoes phase transition with temperature.}
\label{fig4}
\end{figure}

\begin{figure}[h]
\centering
\includegraphics[angle=0.0,origin=c,height=8.8cm,width=8.9cm]{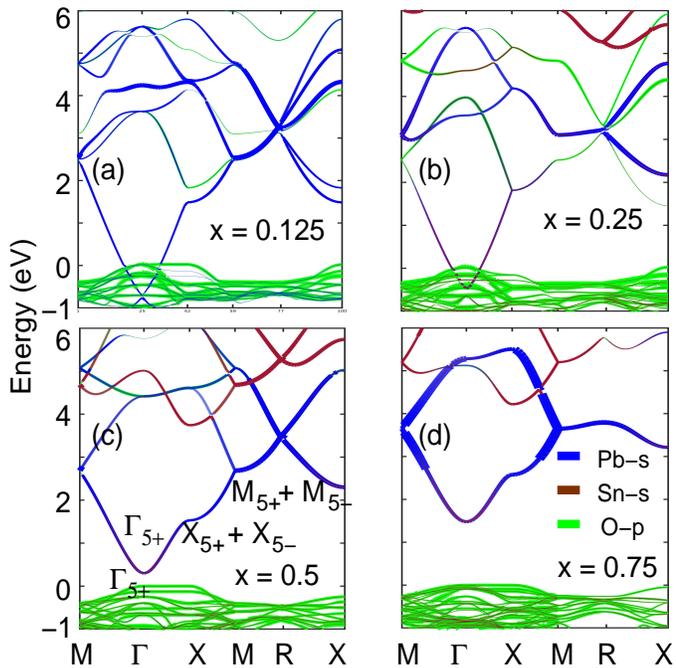}
\caption{ (a-d) Orbital resolved band structure of Sn substituted BPO for various  concentrations. Here, the high symmetry $k$-points correspond to Brillouin zone of 2$\times$2$\times$2 supercell designed for alloys.   }
\label{fig5}
\end{figure}

\begin{figure*}
\centering
\includegraphics[angle=0.0,origin=c,height=7cm,width=16 cm]{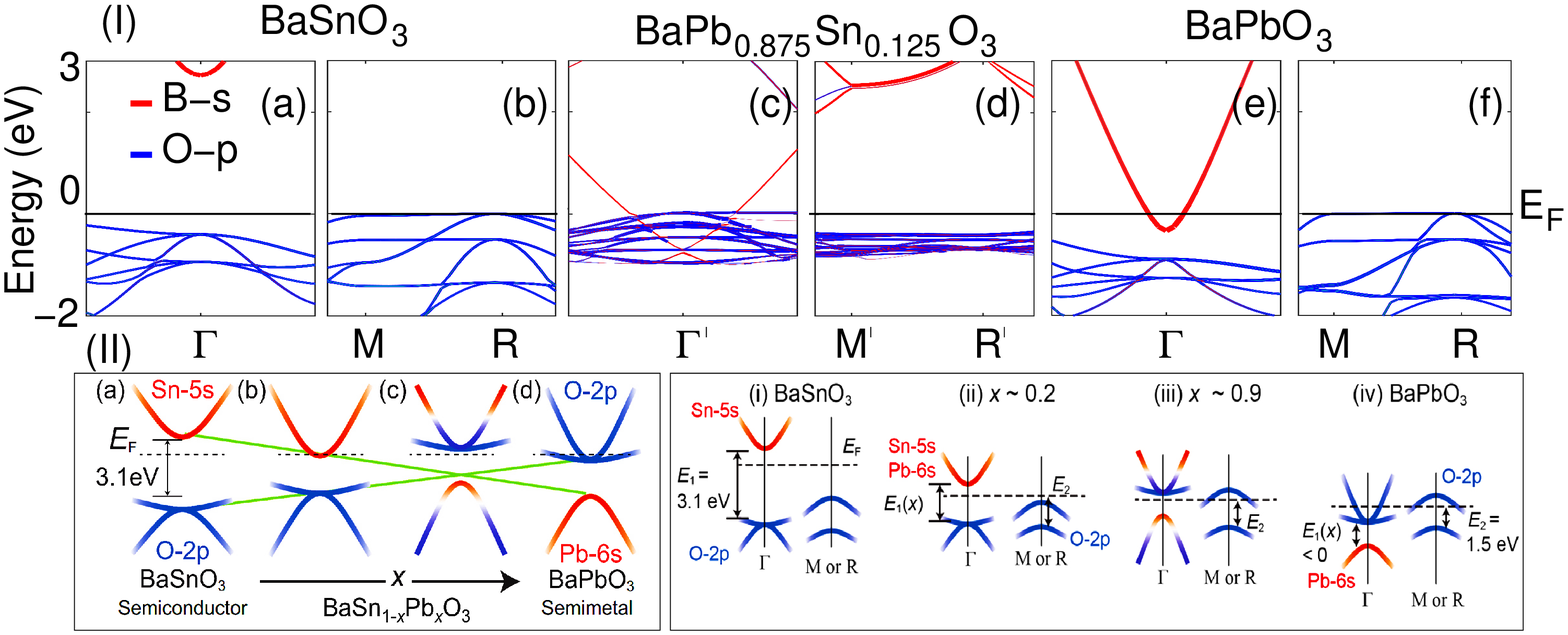}
\caption{  (I) DFT calculated band structure of BSO (a-b), BPO (e-f) and 12.5 \% Sn substituted alloys (c-d) along the selected high symmetry points. While $\Gamma$, M and R are the high symmetry points for the primitive unit cell, $\Gamma'$, M' and R' are that for the supercell adopted to create the alloys.  (II) The proposed band structure for the same in Shiogai et al.\cite{BPO_sn_prb}, which is reproduced with the permission.   }
\label{fig6}
\end{figure*}

 Unlike the case of BPO, experimentally, BSO exists in cubic phase at all temperature range \cite{BaSno_gap, BaSno_struct}. The phonon and electronic band structures are presented in Fig. S5 of the SI. The electronic structure of this phase exhibits insulating phase with an indirect bandgap. Its salient features remains identical to that of BPO (See Fig. \ref{fig1}a). 

 To address our last  objective, we now analyze the electronic structure of Sn substituted BPO to investigate the band topology of alloys. For this, first we considered the cubic phase as the experimental synthesis have indicated the formation of nearly cubic (orthorhombic phase) in absence of any rotation and tilting \cite{BPO_sn_prb}. The band structures of a few representative Sn concentrations in the alloy, with orbital projection, are shown in Fig. \ref{fig5}(a-d). From the figure, we infer the following points. (I) The system remains in the metallic phase for lower values of $x$, as the conduction band still maintains the highly dispersive nature with Pb-s character and creates an electron pocket. The Sn-s dominated band which forms the CB edge in BSO lies far away from the Fermi level in the alloys. (II) As $x$ increases, the width of the conduction band edge decreases. Beyond $x$=0.5, the bandgap opens up gradually. This is due to the homogeneous distribution of Sn, which breaks the Pb-Pb interaction chain. The resulting bandgap and lattice parameter agree with the experimental results provided by Shiogai et al.\cite{BPO_sn_prb}.  However, their proposed band inversion is found to be absent in this study.  To illustrate it further, in Fig. \ref{fig6}, we have compared the band structure of BaPb$_{1-x}$Sn$_x$O$_3$ for $x$ = 0, 0.125 and 1, with the proposed band structure in Shiogai et al. \cite{BPO_sn_prb}. Firstly, we found that Pb-6s and O-2p characters do not form the valence band edge and conduction band edge, respectively, in pure BPO. In fact, it is reverse. The alloy with $x$=0.125 resembles the experimental result of $x$=0.9 (alloyed notation is BaSn$_{1-x}$Pb$_x$O$_3$ in Shiogai et al. and in the current work it is BaPb$_{1-x}$Sn$_x$O$_3$). We find the penetration of the conduction band edge in the valence band spectrum, making it purely metallic.  We also examined the band structure  of alloys in the orthorhombic structure, in which the tilting and rotations are allowed, and observed that the system remains in the insulating phase for all values of $x$, as expected (shown in Fig. S6 of  SI).

In summary, we carried out electronic and phononic band structure calculations with the aid of density functional theory to understand the interplay between lattice and orbital degrees of freedom in Ba(Pb, Sn)O$_3$ and their alloys. The phonon band structure analysis provides insight into three crystal phases in the temperature domain, with the monoclinic phase stabilizing at the lower temperature. The electronic structure analysis indicates that the metallic phase in the cubic BPO is very sensitive to structural distortions and exhibits the metal to insulator phase transition with volume compression, rotation, and tilting of PbO$_6$ octahedra, and with Sn substitution at Pb site. As far as the topology is considered, while no band inversion occurs in the vicinity of the Fermi level to envisage non-trivial electronic phases, the band inversion is in the higher-lying conduction band spectrum observed between Ba+d+Pb-p and Pb-s dominated bands at R point of the Brillouin zone. Sn substitution does not influence the band topology.  

\section*{Supplementary Material}
See the supplementary material for computational details, phonon band structure of all the phases of BaPbO$_3$, and phonon and electron band structure of BaSnO$_3$. 

\section*{ACKNOWLEDGMENTS}
This work was funded by the Department of Science and Technology, India, through Grant No. CRG/2020/004330. We acknowledge the use of the computing resources at HPCE, IIT Madras.

\section*{DATA AVAILABILITY}
The data that support the findings of this study are available
from the corresponding authors upon reasonable request.

\newpage

\begin{center}
    \textbf{\Large Supplementary Material}
\end{center}

\section{Computational details}
In the present work, we have employed both pseudopotential methods with the plane-wave basis set as implemented Vienna Abinitio Simulation Package (VASP), and full-potential linearized augmented plane wave (FP-LAPW) method as implemented in WIEN2k simulation tool \cite{LAPW,Blaha} for DFT calculations.  The pseudopotential method is used for structural optimization, whereas the FP-LAPW is utilized for electronic structure calculations. Perdew-Burke-Ernzerhof (PBE) exchange-correlation functional within the generalized gradient approximation (GGA) is considered in both methods. However, the PBE underestimates the bandgap significantly as compared to the experimental bandgap. Thus, to reduce the error, a modified Becke-Johnson (mBJ) exchange potential along with PBE-GGA correlation is employed \cite{GGA,mbj-1,mbj-2}. The Self Consistent Field (SCF) calculations comprise augmented plane waves of the interstitial region and localized orbitals: B-\{5s, 5p\} (B = Sn, Pb). The R$_{MT}^{I}$ is set to 7.0. The Brillouin zone integration is carried out with a Monkhorst-Pack grid. We used a $k$-mesh of 10$\times$10$\times$10 (yielding 35 irreducible points), 6$\times$6$\times$4 (yielding 35 irreducible points), and 8$\times$8$\times$6 (100 irreducible points) for the cubic, tetragonal and orthorhombic phases, respectively. The phonon band structure is obtained by constructing 2$\times$2$\times$2 supercell and employing the density functional perturbation method as implemented in PHONOPY tool \cite{phononopy}. The Fermi surface is obtained for various lattice parameters of BPO using Wannier based tight-binding model as implemented in Wannier90, and Wanniertools \cite{wannier90,arpes}

\begin{table*}
    \centering
    \caption{ Band degeneracy and irreducible representation of thirteen bands at various high symmetry $k$-points of the cubic Brillouin zone in absence of SOC. Here, Ndg refers to band degeneracy.   }
    \begin{tabular}{cccccccc}
    \hline
    \hline
    &M&&$\Gamma$&&X&&R\\
    Ndg&IR&Ndg&IR&Ndg&IR&Ndg&IR\\
    \hline
1&	A$_{1g}$&	3&	T$_{1u}$&	1&	A$_{1g}$	&1&	A$_{1g}$\\
1&	A$_{2u}$&	3&	T$_{2u}$&	2&	E$_u$	&3&	T$_{2g}$\\
1&	B$_{2g}$&	3&	T$_{1u}$&    1&	A$_{2u}$	&3&	T$_{1g}$\\
2&	E$_g$&	    1&	A$_{1g}$&	1&	B$_{2u}$	&2&	E$_g$\\
2&	E$_u$&	    2&	E$_g$&	    2&	E$_u$	&3&	T$_{1u}$\\
1&	B$_{1g}$&	3&	T$_{2g}$&	2&	E$_g$	&1&	A$_{1g}$\\
1&	A$_{2g}$&&	&		1&	A$_{1g}$	&&	\\
1&	A$_{1g}$&&	&			1&	B$_{2u}$	&&	\\
1&	B$_{2g}$&&	&			1&	A$_{2u}$	&&	\\
1&	A$_{1g}$&&&				1&	B$_{1u}$	&&	\\
2&	E$_g$&	&	&&\\				

\hline
\hline
    \end{tabular}
    \label{T2}
\end{table*}

\begin{figure*}
\centering
\includegraphics[angle=0.0,origin=c,height=5.5cm,width=7.5cm]{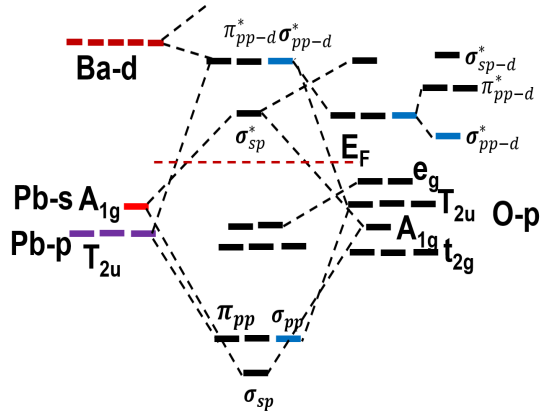}
\caption{ Molecular orbital picture of BaPbO$_3$ constructed from Ba-d, Pb-\{s,p\} and O-p atomic orbitals. These atomic orbitals undergo symmetry adapted linear combination with respect to point group symmetry and forms the new set of basis denoted them in irreducible representation form. Here, the molecular orbital is presented for O$_h$ point group symmetry exists at R and $\Gamma$ high symmetry points.  }
\label{fig1}
\end{figure*}

\begin{figure*}
\centering
\includegraphics[angle=0.0,origin=c,height=6.0cm,width=9cm]{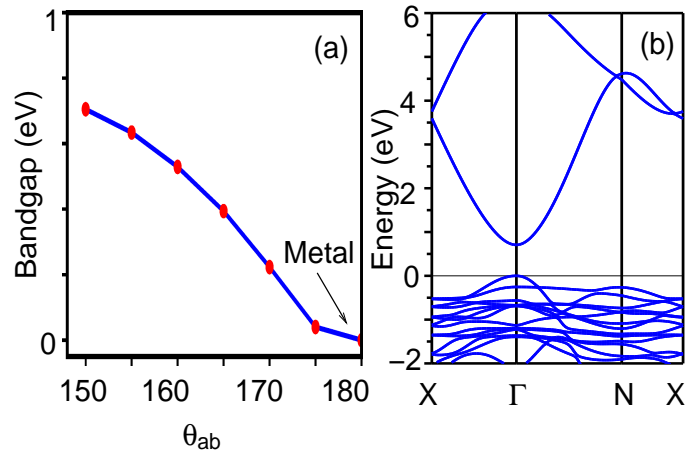}
\caption{ Variation of electronic bandgap as a function of in-plane octhahedral rotations for constant unitcell volume. (b) Band structure of tetragonal phase  for in-plane octahedral angle $\theta_{ab}$=150$^{\circ}$. }
\label{fig2}
\end{figure*}

\begin{figure*}
\centering
\includegraphics[angle=0.0,origin=c,height=7.3cm,width=9.5cm]{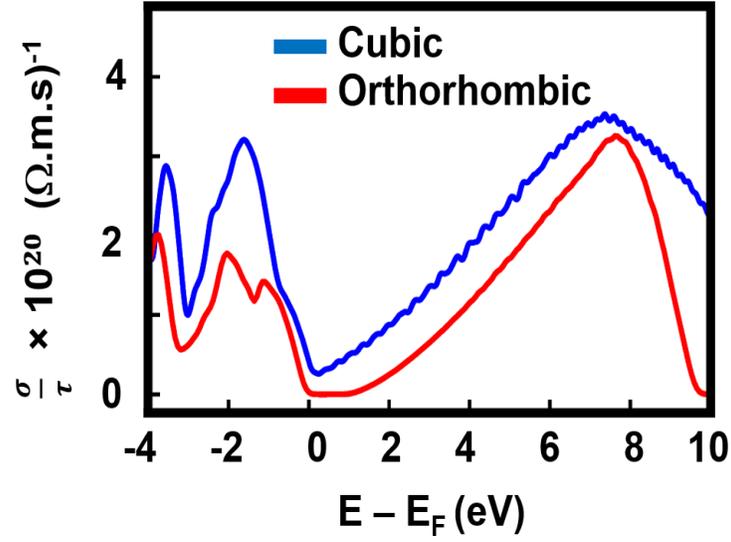}
\caption{Trace of the electrical conductivity of cubic and orthorhombic phases.   }
\label{fig3}
\end{figure*}

\begin{figure*}
\centering
\includegraphics[angle=0.0,origin=c,height=12.0 cm,width=17.6cm]{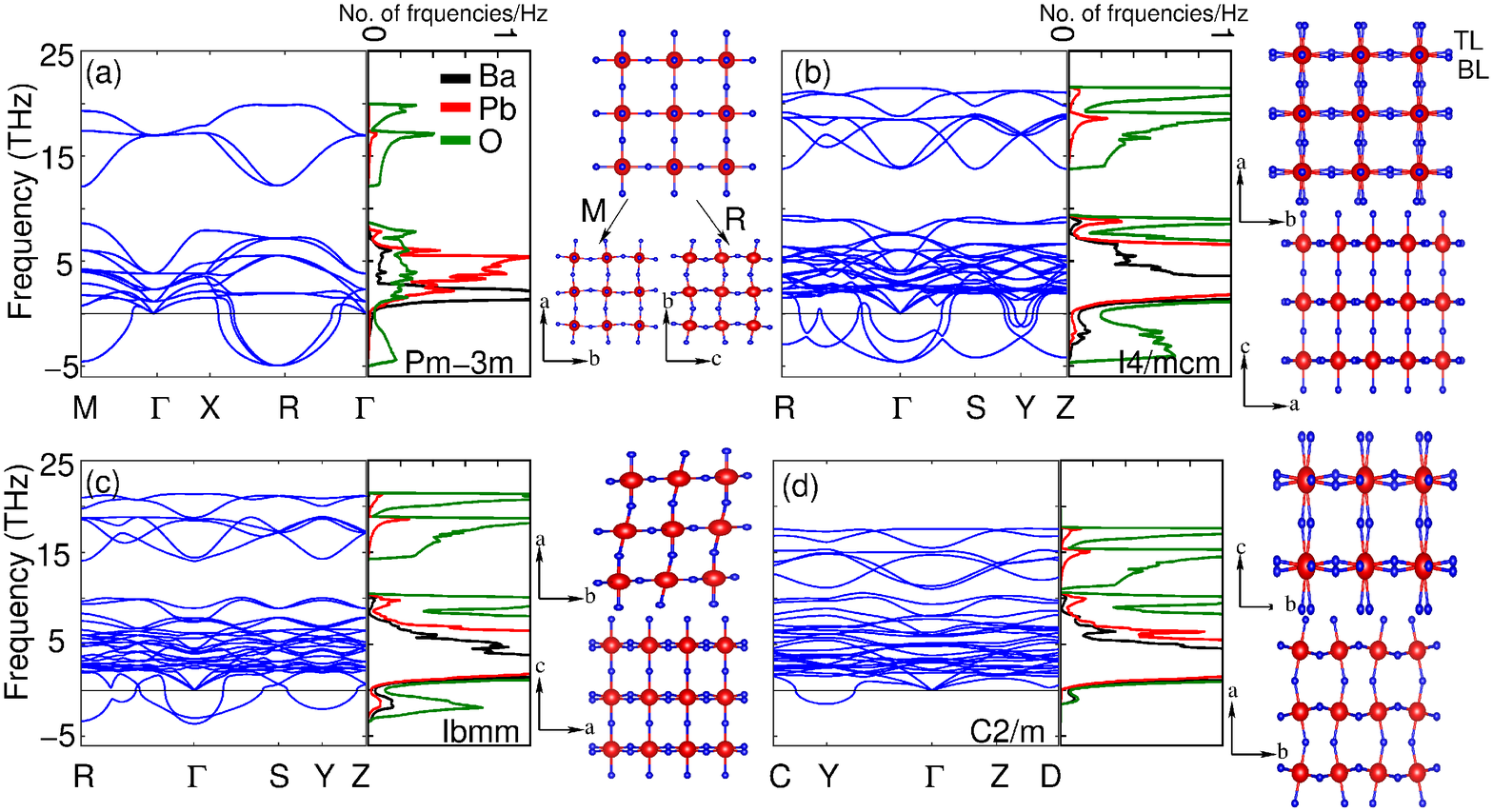}
\caption{Phonon band dispersion for cubic (a), tetragonal (b), orthorhombic (c) and monoclinic (d) BPO obtained from density functional perturbation theory. The soft modes in the cubic phase band dispersions infer that the structure undergoes phase transition with temperature.}
\label{fig4}
\end{figure*}

 \begin{figure*}
\centering
\includegraphics[angle=0.0,origin=c,height=11.5cm,width=10.0cm]{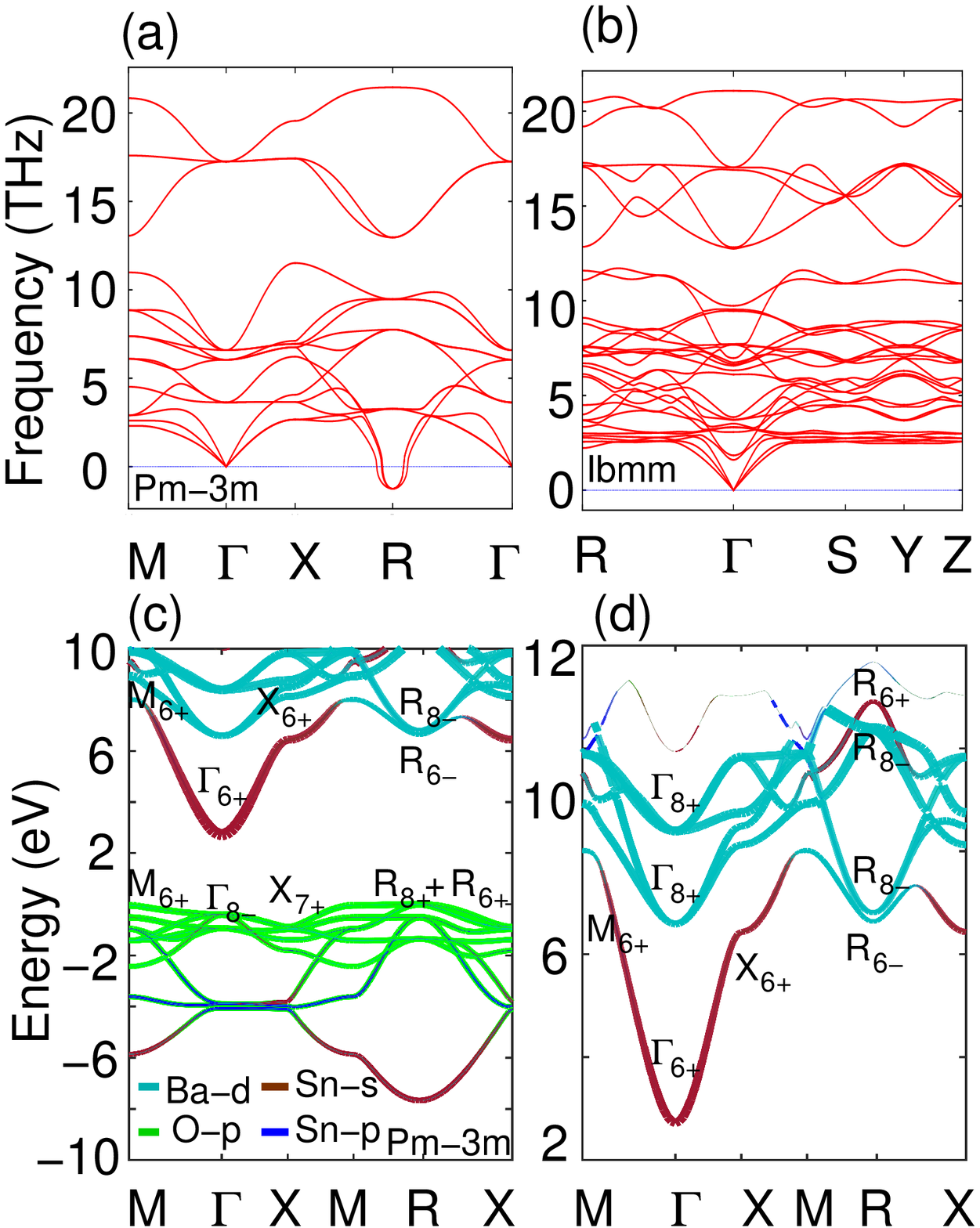}
\caption{ (a-b) Phonon band structure of cubic and orthogonal BSO. Presence of soft modes in the cubic phase indicates the instability at 0 K. The orthorhombic phonon structure infers the stable phase in low temperature regime. (c) Orbital resolved electronic band structure for cubic phase and (d) conduction band spectrum showing the band inversion at R point.     }
\label{fig5}
\end{figure*}

 The phonon band structure shown in the Fig. \ref{fig5}a of cubic BSO demonstrates soft mode at the R point of the Brillouin zone. Thus there exists a possibility of minor lattice distortion. Keeping this in mind, we carried out complete structural relaxation, which yields orthorhombic structure with space group Ibmm. The corresponding phonon band structure is shown in Fig. \ref{fig5}b, which shows the absence of soft modes. The total energy of the cubic phase is found to be just 14 meV/f.u. which is higher than that of the orthorhombic phase. This suggests that the experimentally synthesized high-temperature phase is likely to be cubic. The electronic band structure for cubic phase is shown in Fig. \ref{fig5}c-d. The system is in the insulating phase and exhibiting an indirect bandgap in the cubic phase. When compared to the band structure of BPO (see Fig. \ref{fig2}a), the salient features are nearly identical except the fact that, in the case of BPO, the bandgap vanishes with the formation of electron pocket at $\Gamma$, as already been discussed.  In the case of BSO, VBM and CBM exhibit even parity, which results in parity forbidden transitions. However, at $\Gamma$ point, the valence band edge has odd parity, and thus transitions can occur with a less intense absorption peak. Hence, the optical bandgap is higher than the electronic bandgap \cite{BaSno_gap}.

\begin{figure*}
\centering
\includegraphics[angle=0.0,origin=c,height=6.0cm,width=12.5cm]{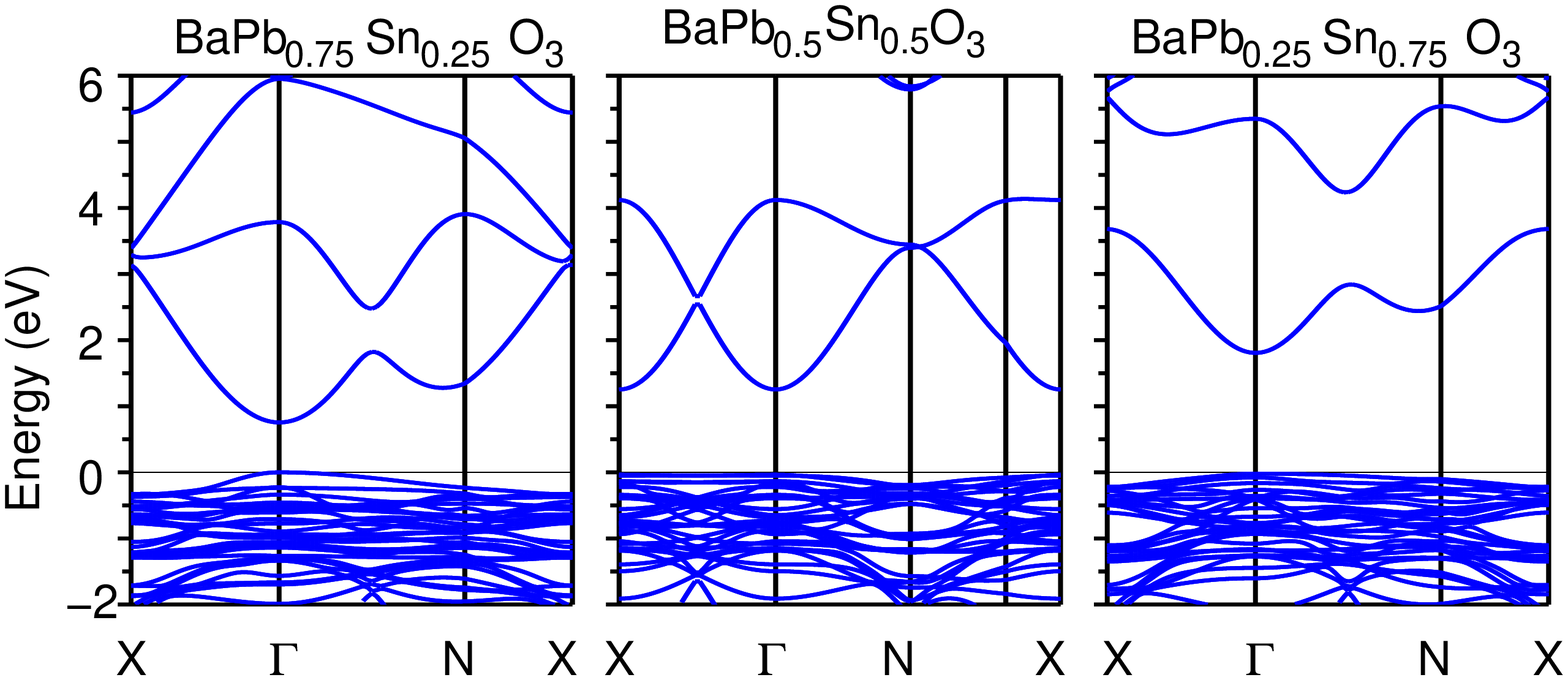}
\caption{ Band structure of Sn substituted Pb alloyed perovskites for the orthorhombic phase.   }
\label{fig6}
\end{figure*}
\newpage
\section*{REFERENCES}
\bibliography{paper}

\end{document}